%% file: mgr.tex
\documentstyle[11pt,emulateapj]{article}
\input psfig.tex
\begin{document}
\title{Faint field galaxies: an explanation of the faint blue excess
using number evolution models}
\author{Ping He$^{1,2}$, Zhen-Long Zou$^{1,2}$, and 
  Yuan-Zhong Zhang$^{2,3}$}
\affil{$^1$Beijing Astronomical Observatory, National Astronomical
Observatory, Chinese Academy of Sciences, Beijing 100012, P.R. China, 
 and CAS-PKU Joint Beijing Astrophysics Center, Beijing 100871, P.R. 
China (hemm@itp.ac.cn)}
\affil{$^2$Institute of Theoretical Physics, Academia Sinica, P.O.Box
 2735, Beijing 100080, P.R.China}
\smallskip
\affil{$^3$CCAST (World Laboratory), P.O.Box 8730, Beijing 100080, 
P.R.China}

\begin{abstract}

Pure luminosity evolution models for galaxies provide an unacceptable fit to the
redshifts and colors of faint galaxies. In this paper we demonstrate, using HST
morphological number counts derived both from the $I_{814}$-band of WFPC2
in the Medium Deep Survey (MDS) and the Hubble Deep Field (HDF) and from the
$H_{1.6}$-band of NICMOS, and ground-based spectroscopic data of the Hawaii
Deep Field and the Canada-France Redshift Survey, that number evolution is
necessary for galaxies, regardless of whether the cosmic geometry is flat, open, or
$\Lambda$-dominated. Furthermore, we show that the  number evolution is small
at redshifts of $z<1$, but large at $z>1$, and that this conclusion is valid for all the
three cosmological models under consideration. If the universe is open or
$\Lambda$-dominated, the models, which are subject to the constraint of the
conservation of the comoving mass density of galaxies, naturally predict a population
of star-forming galaxies with the redshift distribution peaking at $z=2\sim3$, which
seems to be consistent with the recent findings from Lyman-break photometric
selection techniques. If the cosmological model is flat, however, the conservation of
the comoving mass density is invalid. Hence, in order to account for the steep slope
of $B$-band number counts at faint magnitudes in the flat universe, such a
star-forming galaxy population has to be introduced ad hoc into the modelling
alongside the merger assumption.

\end{abstract}

\section{Introduction}

The origin and nature of the excess population of faint blue galaxies (FBG) 
has been a topic of intense debate for many years (Koo \& Kron 1992; Ellis 
1997). The number counts of galaxies at blue ($B$, $\lambda_{eff}$= 4500
\AA) and near-infrared ($K$, $\lambda_{eff}$=2.2 $\mu$m) 
wavelengths produce conflicting results. The $B$-band counts show an excess 
over the no-evolution predictions, and suggest strong luminosity evolution in 
galaxy populations, while the $K$-band counts are well fitted by models with
only passive evolution. When spectroscopic samples are available, it is 
found that the redshift distributions of galaxies are also consistent with 
passive evolution while strong luminosity evolution will lead to a 
high-$z$ distribution, in excess of the observations. To avoid these 
paradoxes, a number of scenarios have been suggested. The first is pure 
luminosity evolution (PLE) in galaxy populations, which increases the distance 
to which galaxies may be seen (Tinsley 1980; Bruzual \& Kron 1980; Koo 1981, 
1985; Guiderdoni \& Rocca-Volmerange 1990; Gronwall \& Koo 1995; Pozzetti, 
Bruzual, \& Zamorani 1996). The second attempt is the choice of a cosmological
geometry that maximizes the available volume, either by adopting a low value of 
the deceleration parameter $q_0$ (an open cosmological model) or by introducing
a cosmological constant $\lambda_0$ (Broadhurst, Ellis \& Shanks 1988; Colless 
et al. 1990; Cowie, Songaila \& Hu 1991; Colless et al. 1993; Fukugita et al. 
1990). The third is to increase the number of galaxies at earlier times, either by 
introducing additional populations, which once existed at high $z$ but have 
since disappeared or self-destructed (Broadhurst, Ellis, \& Shanks 1988; Cowie 
1991; Babul \& Rees 1992; Babul \& Ferguson 1996; Ferguson \& Babul 1998), or 
by merging, that is by assuming that present-day galaxies were in smaller 
fragments at high redshifts (Rocca-Volmerange \& Guiderdoni 1990; Cowie et al. 
1991; Guiderdoni \& Rocca-Volmerange 1991; Broadhurst, Ellis \& Glazebrook 
1992; Carlberg \& Charlot 1992; Kauffmann, Guiderdoni \& White 1994; Roukema 
et al. 1997).

One constraint on the nature and evolutionary behavior of faint galaxies arises 
from the morphological number counts derived from the Hubble Space Telescope 
(HST), whose unprecedented imaging ability enables galaxies to be segregated 
morphologically into several broad classes (Glazebrook et al. 1995a; Driver et 
al. 1995; Abraham et al. 1996; Teplitz et al. 1998). These morphological data 
have provided us with valuable information to understand the origin and 
evolution of galaxies. Now, any self-consistent theory of galaxy evolution and 
cosmology must pass the test of matching the distribution of galaxy properties 
derived from HST morphological data.

In a previous investigation (He \& Zhang 1998, hereafter HZ98), we modelled
the number counts of E/S0 galaxies obtained from the MDS and the HDF in the
HST $I_{814}$-bandpass ($\lambda_{eff}$=8000\AA), and found that the 
number counts of ellipticals could be well explained by PLE models in any 
cosmological model if ellipticals are assumed to have formed at high redshifts
(say $z_f$=5.0) and thereafter have passively evolved (i.e., no further star
formation). This is just the traditional scenario of the formation and evolution
for elliptical galaxies. However, incorporating other observations such as redshift
and color distributions and the modelling of the other types (i.e., early- and
late-type spirals and irregulars, as separated by HST), we have found that the
FBG problem cannot be resolved by the conventional scenario for elliptical
galaxies, and the inclusion of dust extinction also cannot save the scenario. This
conclusion holds in each of the three cosmological models under consideration:
flat, open and $\Lambda$-dominated. Our investigation suggests that number
evolution  may be essential to explain the faint data (He \& Zhang 1999a,
hereafter HZ99a).

Number evolution models have been widely considered to account for the 
paradoxes concerning the FBG problem, whereas PLE model is the starting
point of our investigation. We model the number counts of the three classes
of galaxies- ellipticals, early-type spirals, and late-type spirals/irregulars-
which are derived from the HST $I_{814}$-passband, and in combination
with the analysis of redshift distributions in the observed magnitude bins, we
find that PLE models can explain the number counts at the bright magnitudes
of $I_{814}<22.5$. This indicates that PLE models could be valid up to $z\sim1$,
in apparent contradiction to scenarios of strong number evolution at low redshifts
(e.g., Kauffmann, Charlot, \& White 1996). Kauffmann et al. concluded that only
one third the number of nearby bright elliptical galaxies were in place at $z\sim1$,
which would seem to be inconsistent with the number counts at the bright end.
Our conclusion is consistent with those of Roukema \& Yoshii (1993) and Woods
\& Fahlman (1997) based on the analysis of angular correlation function, and
that of Totani \& Yoshii (1998) based on the $V/V_{max}$ analysis (Schmidt
1968), and in particular, agrees well with the results of Im et al. (1999), based
on the morphologically divided redshift distribution of faint galaxies, and those
of Brinchmann et al. (1998) and Lilly et al. (1998), based on the HST imaging of
the CFRS and LDSS redshift surveys.

We tentatively construct a set of phenomenological number-luminosity evolution 
(NLE) models to explain data such as the morphological number counts in both 
the $I_{814}$-band of the Wide Field and Planetary Camera 2 (WFPC2) and the 
$H_{1.6}$-band of the Near-Infrared Camera and Multiobject Spectrometer 
(NICMOS), as well as the counts in $B$- and $K$-bands derived from most of the 
ground-based observations and redshift distributions derived from the Hawaii 
sample (Cowie et al. 1996) and the Canada-France Redshift Survey (CFRS) (Lilly 
et al. 1995a). The significant feature of these models is that the number evolution
is small at redshifts of $z<1$, while it is large at high redshifts. We find that NLE
models can roughly reproduce these observations.

To simplify the modelling, we employ the constraint of conservation of the
comoving mass density of galaxies, in the form of Guiderdoni \& Rocca-Volmerange
(1991). We find that a population of late-type spiral/irregular galaxies and
star-forming galaxies naturally emerge in our models, due to mass conservation.
Such a population will peak at $z\approx2.8-3$. Between $z\sim3$ and the present
day, it behaves as a population which seems to be fading, disappearing or
self-destructing, with very blue colours. Such a population also seems to be consistent
with the recent findings using Lyman-break selection techniques (see Steidel \& 
Hamilton 1992, 1993; Steidel, Pettini, \& Hamilton 1995; Madau 1995). This 
conclusion, however, is only valid in open (e.g., $\Omega_0$=0.1) or 
$\Lambda$-dominated (e.g., $\Omega_0$=0.2, $\lambda_0$=0.8) cosmological 
models. If the universe is flat ($\Omega_0$=1), the comoving mass density of 
such a predicted population becomes negative at high redshifts ($z>3$), which 
is physically unreasonable. Therefore, to account for the faint data, we abandon
the constraint of mass conservation and introduce a star-forming population
{\it ad hoc} into the modelling. 

In Section 2, we present the PLE models. We are particularly interested in the behaviors
of the models at bright magnitudes, or at low redshifts. Section 3 is devoted to the
number evolution models. The summary and conclusions are presented in Section 4.

\section{Pure Luminosity Evolution}

PLE model is the starting point of the investigation. This kind of model takes as
ingredients the stellar evolutionary tracks, normally for a restricted metallicity
range, the initial mass function (IMF), and a gas consumption timescale associated
with star formation rate (SFR), which are adjusted to give the present range of
colors across the Hubble sequence. With each galaxy assumed to evolve as an
isolated system, the rest-frame spectral energy distribution (SED) can be predicted
at a given time and thus an  evolutionary correction can be determined (Ellis 1997).
The evolutionary SEDs of galaxies are computed using population synthesis models,
(e.g.,  the latest Bruzual \& Charlot synthesis code, which we employ in this work). 
The reader is referred to HZ98, HZ99a or Pozzetti et al. (1996) for details of the
modelling. We also consider the effects of dust extinction, following Wang's (1991)
prescription, which is similar to that of HZ98 and HZ99a. However we choose a larger
optical depth of $\tau^{*}=0.3$ than that adopted by Wang for galaxies other than
ellipticals (see also Campos \& Shanks 1997).

Similar to the prescription of HZ98 and HZ99a, we adopt three representative 
cosmological models in this work: 1) flat, $\Omega_0=1.0$, $\lambda_0=0$, and 
$h=0.5$ ($H_0$=100 $h$ km s$^{-1}$ Mpc$^{-1}$) (hereafter, Scenario A),
2) open, $\Omega_0=0.1$, $\lambda_0=0$, and $h=0.5$ (Scenario B), and 3)
flat and $\Lambda$-dominated, $\Omega_0=0.2$, $\lambda_0=0.8$, and $h=0.6$
(Scenario C). We assume a formation redshift $z_f=5.0$ for all types of galaxies in
all three cosmological models.

We first consider the modelling of the morphological number counts derived 
from the MDS and the HDF in the $I_{814}$-passband. The basic ingredient
of the models, the present-day luminosity function (LF), is not well-determined
by local surveys due to inherent uncertainties such as a large local fluctuation
(Shanks 1989), significant local evolution (Maddox et al. 1990a), selection effects
and/or incompleteness (Zwicky 1957; Disney 1976; Ferguson \& McGaugh 1995),
or perhaps systematic errors in local surveys (Metcalfe, Fong \& Shanks 1995a).
Therefore, a simple 3 parameter Schechter function ($\phi^*$, $M^*$ and
$\alpha$, cf., Schechter 1976), as adopted by most authors, seems not to be
universal, and hence not to be quite appropriate as an average over all types and
environments (cf., Binggeli, Sandage, \& Tammann 1988). In particular, much recent
work finds a tail for dwarfs and late types that is steeper than $\alpha\sim$-1.3
(Driver, Windhorst \& Griffiths 1995), and in addition, low surface brightness galaxies
(LSBGs) may make up a significant amount of the luminosity density of the local
universe (see Impey \& Bothun 1997 for a review about LSBGs). Regardless of these
uncertainties, the simple Schechter form for local LFs is still our working assumption
in the current research.

Firstly, we adjust the Schechter LF parameters by trial and error so that the
model predictions can match better the bright-end number counts, since the count
normalization is sensitive to the three parameters $\phi^*$, $M^*$ and $\alpha$,
especially the former two (cf. Ellis 1997). The LFs for galaxies are listed in Table 1.
Secondly, we test these ad hoc adoptions by comparison with observations. It can be
seen, in Figure 1, that both the $B$- and the $K$- band adopted LFs lie close to the
observations, and in particular, the observed faint-end slopes are not as high as, say
-1.5, as predicted by the semi-analytic models (cf. Cole et al. 1994; Kauffmann,
White \& Guiderdoni 1993).

\subsection{Number counts for elliptical galaxies}

We adopt the Scalo (1986) IMF for ellipticals and the Salpeter (1955) IMF for the
other types. As addressed in HZ98, the Scalo IMF is less rich in massive stars than
the Salpeter IMF because of the steeper slope of the former at the high-mass end.
The SFR for ellipticals is adopted as the simple exponential form
$$\psi(t)\sim {\rm exp}(-t/\tau_e), \eqno(1)$$ where $\psi(t)$ refers to the SFR,
and $\tau_e$ is the timescale characterizing this kind of SFR. A single value of
$\tau_e$ can not satisfactorily reproduce the $B-K$ color distribution (He \& Zhang
1999b), which may be a reflection of intrinsic complexities such as the metallicity
variations among the population of ellipticals. Besides, there is also an age-metallicity
degeneracy (Worthey 1994), which greatly complicates the situation. For the sake of 
simplicity, we neglect the chemical evolution in the current work. In our models,
ellipticals are subdivided into five parts equal in number, with each one of 
them specified by a different $\tau_e$ for SFR to characterize its luminosity 
evolution. By such treatments, we hope to circumvent the metallicity-age 
degeneracy. The values of $\tau_e$, modelled $B-I$ and $B-K$ colors are shown 
in Table 2. Notice that the observed average $B$-$K$ color for present-day 
ellipticals is $\sim4.10$ (Johnson photometric system), with the dispersion 
being about $0.1-0.2$ mag, and hence these values of the parameter $\tau_e$ 
are acceptable. It is well known that the local photometric and spectroscopic
properties of ellipticals are largely insensitive to the past star formation history
(cf. Bruzual 1983; Bower, Lucey \& Ellis 1992; Charlot \& Silk 1994).

We show in Fig.2-(a) the predicted number counts for ellipticals by PLE models 
to compare with those derived from HST in $I_{814}$-passband. It can be seen 
that up to $I_{814}$=22.5, the predictions can satisfactorily reproduce the 
bright-end number counts in all the three scenarios, indicating the PLE models 
are valid within this magnitude range. At the faint-end ($I_{814}>22.5$), 
however, the predictions in Scenarios B and C largely overestimate the 
observations. Although the prediction for Scenario A fits the bright counts well,
as we have analyzed in HZ99a, models with strong luminosity evolution 
predict too many high-$z$ objects, in substantial disagreement with the 
color-selected sample from Cowie et al. (1996).

Though PLE models are unacceptable, we present redshift distributions for
ellipticals in the observed $I_{814}$ magnitude bins, in order to better
understand the models. In Fig.2-(b), (c) and (d), we show predictions of
redshift distributions in the three magnitude bins, $I_{814}<21.0$, 
$21.0<I_{814}<22.5$, and $22.5<I_{814}<25.0$. The redshift distributions
peak around $z\sim0.5$, indicating the counts at $I_{814}<21.0$ are
contributed by the objects at the redshifts of $z<0.5$ or even as large as
$z\sim0.8$ (the half maximum of the peaks), and from Fig.2-(c), it can be seen
that the peaks appear at $z\sim0.8$, indicating the objects between
$21.0<I_{814}<22.5$ are mostly at $z<0.8$ or $\sim1.0$. From the previous
analysis, we know that the number counts at $I_{814}<22.5$ can be well
explained by PLE models up to $z\sim1$. As addressed above, the predicted
number counts at $I_{814}>22.5$ in Scenarios B and C significantly exceed the
observations, and from Fig.2-(d), we know that these excessive objects mostly
lie in $z \geq 1$. This implies, for example, that if a merger scenario for the
evolution of ellipticals is involved, then the merger rate in the interval $0<z<1$
seems to be small. This conclusion is valid in all the three world models under
consideration.

We now consider an example of strong number evolution for ellipticals. Kauffmann,
Charlot \& White (1996) claimed, based on $<V/V_{max}>$ analysis (Schmidt 1968)
and the data from the CFRS and the Hawaii Deep Survey (Cowie et al. 1996), that
only one third of the nearby bright elliptical galaxies existed at $z\sim1$. We adopt
their expression of the number density evolution for ellipticals with respect to $z$, i.e., 
$F=(1+z)^{-\gamma}$, with $\gamma=1.5\pm0.4$ (This means $F\approx 1/3$
at $z=1$), to compare the predictions with observations. We show the model 
predictions in Fig.2-(a). Obviously, such a strong number evolution cannot be accepted,
since it underestimates the bright-end number counts. Notice that the bright-end counts
are mostly contributed by  $L^*$ galaxies, and hence they are insensitive to the LF
faint-end slope $\alpha$. An enhancement of the normalization $\phi^*$ by, say
50\% can improve the goodness of fit, but such an enhancement will produce too many
bright ellipticals at the present day, and conflict with the local surveys, especially the LF
at $K$-band. It is worthy of mentioning that Totani \& Yoshii (1998) have performed a
detailed $V/V_{max}$ test, and concluded that the value of $<V/V_{max}>$ for
ellipticals is $\sim$0.5 in the redshift range $0.3<z<0.8$, which means that the number
density is nearly unchanged in this redshift range, and this conclusion is similar to ours.
We conclude that strong number evolution for ellipticals at low redshifts is not acceptable.

\subsection{Number counts for early-type spiral galaxies}

Using the morphological classification of galaxies by HST, we combine Sab and Sbc
galaxies into a group of early-type spirals. The LFs are presented in Table 1. The SFR
is taken the same exponential form as Eq. [1], but the timescales are generally larger
than those of ellipticals in order to match their local photometric properties. These
parameters are listed in Table 2. 

In Fig.3-(a), we present our PLE model predictions of number counts to compare with
the observations. The models satisfactorily reproduce the observations at almost all
magnitudes in Scenarios B and C, indicating little number evolution. For Scenario A,
however, the PLE model cannot account for observations at $I_{814}>22.5$, and hence
other assumptions than PLE are needed. As is Section 2.1, we analyze the redshift
distributions in the observed magnitude bins.

We can see from Fig.3-(b) and (c) that, in the magnitude bins $I_{814}<21.0$ and
$21.0<I_{814}<22.5$, there are peaks in the redshift distributions at
$z\approx0.5\sim0.8$, indicating that PLE models could be valid even out to the redshift
of the half maximum of the peaks, i.e., $z\approx0.8\sim1.0$. For $I_{814}>22.5$,  we
can see from Fig.3-(d) that the models fail to predict enough early-type spirals at the
redshifts of $z>1.0$.

\subsection{Number counts for late-type spiral/irregular galaxies}

We mix Scd and Sdm/Irr classes into a category of late-type spiral/irregular galaxies.
In addition, we introduce another population which has very blue colors (even bluer 
than irregular galaxies), and we refer to it as the vB galaxies with the name  taken after
Pozzetti et al. (1996). The vB galaxies are included to explain the bluest colours,
$B-K\approx2$ of the Hawaii sample and  $(V-I)_{AB}\approx0.5$ of the CFRS sample.
As addressed by Totani \&  Yoshii (1998), such very blue galaxies comprise nearly 10\% of
the CFRS sample. The very blue colors suggest that they are star-forming galaxies, and
hence vB galaxies are meant to reproduce the population of star-burst galaxies present 
at each $z$, whose evolution does not follow the prescription of standard population
synthesis. Following the treatment of Pozzetti et al., we assume that star formation in
these galaxies keeps their SEDs constant in time, and the SEDs can be approximated by a
model with constant SFR lasting 1.5 Gyr.

Their LFs are also listed in Table 1. In Fig.4-(a), we show the predicted 
number counts for comparison with observations. Similar to the cases of 
ellipticals and early-type spirals, the PLE models can also reproduce the 
late-type spiral/irregular population to $I_{814}<22.5$, indicating that PLE 
models could be valid in the redshift range $z<0.8\sim1$ (Fig.4-(b) and (c)), 
while at $I_{814}>22.5$, especially for Scenario A, the models underpredict 
the observations, indicating the models need to be improved at high redshifts 
of $z>1$ (Fig.4-(d)).

In short, PLE models can well explain the number counts of all types of 
galaxies at the magnitudes of $I_{814}<22.5$, or equivalently, according to 
the analysis of redshift distributions, at the redshifts of $z<1$. Hence,
strong number evolution at low redshifts is not acceptable, comparing to the 
observations.

\section{Number-Luminosity evolution}

Pure luminosity evolution works fairly well at low redshifts, $z<1$, as suggested
by our previous investigations. Next, without considering the underlying physical
mechanism, we phenomenologically construct a set of number evolution models to
explain the faint data. There is no doubt that the luminosity evolution must occur,
due to the aging of star populations. Moreover, if mergers between galaxies are
included in the modelling, it is more realistic to consider the additional star formation,
that can be induced by interactions between galaxies. For simplicity, we use the same
galaxy evolution models as in Section 2 to compute the $K$-corrections and
evolutionary-corrections for galaxies,  without considering more complicated star
formation scenarios (e.g., Col\'in,  Schramm, \& Peimbert 1994; Fritze-v.Alvensleben
\& Gerhard 1994), or various complicated physical processes concerning the formation
of galaxies (cf. Roukema et al. 1997).

Number evolution can be simply realized by making the LF parameters $\phi^*$,
$M^*$ and $\alpha$ functions of $z$. For this, we adopt the following expressions: 
$$\phi^*(z)=\phi^*{_0}(1+z)^{Q_0+Q_{2}z^2}, \eqno(2)$$
where $\phi^*_0$ refers to the characteristic density of present-day galaxies.
$Q_0$ and $Q_2$ are two free parameters. This form is a refinement of our
previous formalism (cf. HZ99a), and can also be treated as a modification of that
of Guiderdoni \& Rocca-Volmerange (1991), in which the power-law index is
not a function of $z$. Obviously, $Q_2$  will mostly affect the high-$z$ behavior.
The evolution of the faint-end slope  $\alpha$ is parametrized as the following:
$$\alpha(z)=\alpha_0-\gamma{_1}\left(
{z\over z{_f}}\right)-\gamma{_2}\left({z\over z{_f}}\right)^2, 
\eqno(3)$$
where $\alpha_0$ represents the local quantity, and $z_f$ is the formation redshift.
The two parameters $\gamma{_1}$ and $\gamma{_2}$ characterize the functional
relationship between $\alpha$ and $z$. In principle, $M^*$ should also be a
function of $z$. We find, however, the variation of $M^*$ can be to some extent
compensated by that of $\phi^*$ and $\alpha$, and for simplicity, we neglect the
variation of the characteristic magnitude. Our model is therefore similar to the 
$\phi^*$-$\alpha$ model of Guiderdoni \& Rocca-Volmerange (1991).

Thus we have four free parameters, $Q_0$, $Q_2$, $\gamma_1$ and $\gamma_2$,
with which to characterize the number evolution. Nevertheless, for a specific Hubble
type, there is no need to invoke simultaneously all the four parameters. Our strategy
to proceed is to involve as few of these parameters as possible, as long as the models
can reproduce the observed data. We model the three Hubble types independently.

\subsection{Morphological number counts}

We first consider the explanation of elliptical number counts by the NLE models. From
Section 2, we realize that the number evolution is small at $z<1$, hence the (absolute)
value of the parameter $Q_0$ should be low. Since the PLE predictions exceed the 
observations at faint magnitudes (i.e, at high redshifts), the parameter $Q_2$ should be
negative. We try to fix these two parameters by trial and error to match the data. It can
be seen from Panel (a) of Figure 5, that the NLE models can reproduce well the data at
both bright and faint magnitudes in each of the three Scenarios. The values of $Q_0$ and
$Q_2$ are listed in Table 3, and  in Figure 6, we show the relationship of $\phi^*$ with
$z$. We can see that  the number density of ellipticals is nearly unchanged in the redshift
range of $z<1$; while at high redshifts, ellipticals are absent regardless of the 
cosmological models. This conclusion agrees well with the finding of Totani \& Yoshii
(1998) based on the $V/V_{max}$ analysis, and is also in agreement with that of Zepf
(1997).

The number counts of early-type spirals can also be satisfactorily reproduced by our
models (See Panel (b) of Figure 5). For Scenarios B and C, only $\gamma_1$ and
$\gamma_2$ are needed, and $Q_0$ or $Q_2$ is not necessary. There are additional
advantages for such $\alpha$-variation models, that the faint-end counts are
contributed by dwarfs rather than bright galaxies, and hence there are no high-$z$ tails
predicted by the models. For Scenario A, it seems that $Q_2$ is also needed, since the flat
universe has less volume at high redshifts than the open or $\Lambda$-dominated
universe, and hence a slightly positive $Q_2$ can help to improve the fit at faint
magnitudes. The values of these parameters are also listed in Table 3. We can see that
the variation of $\alpha$ at low redshifts is also not dramatically large. For example,
$\alpha$ steepens only by 0.06 from $z=0$ to $z=1$ in Scenario A. 

The modelling for the late-type spiral/irregular and vB galaxies is similar to that for
ellipticals and early-type spirals, but requires even more free parameters. As a
result, we proceed in an alternative way. In a scenario of hierarchical clustering,
galaxies are constantly merging over cosmic time. If the hierarchical picture is realistic,
and in the absence of other scenarios such as fading or self-destructing of a galaxy
population, then the total comoving masses of gas and stars can be regarded to be
enclosed within galaxies. Hence the total comoving mass density should be conserved in
merger processes. We employ the conservation of mass to constrain the number density
evolution of the latest Hubble types.

Mass (baryonic, including gas and stars) conservation can be expressed in the following
way: 
    $$\sum\limits_{j}M_j^*(z)\phi_j^*(z)\Gamma(\alpha_j(z)+2)={\rm constant},
    \eqno(4)$$ 
which is taken from Guiderdoni \& Rocca-Volmerange (1991). $M^*$ is the 
characteristic mass of the galaxy mass function (MF). The total mass density is
computed by summing over galaxy type $j$, and the constant can be determined by
present-day quantities. In order to obtain the evolution of LFs, we need to convert the
evolving MFs into evolving LFs, and hence the mass-to-luminosity ($M/L$) ratios are
needed. Unfortunately, the $M/L$ ratios are poorly determined, especially for
$B$-passband. However, the near-infrared light is less affected by star formation,
and thus by evolution, than the optical emission, and it is a good tracer of the total
mass within galaxies. Hence we will make use of $K$-band LFs instead of the MFs of 
galaxies.

We assume the faint-end slope $\alpha$ for late-type galaxies (Scd, Sdm/Irr, and vB)
evolves in the same way as for the early-type spirals (Sab and Sbc). In this case the
evolution of $\phi^*$ can be determined by Eq. [4]. The evolution of the characteristic
number density $\phi^*$ for $L^*$ galaxies is illustrated in Figure 7. We first analyze
the cases for Scenarios B and C. Using the constraint of mass conservation suggested
by Eq. [4],  $\phi^*$ increases with increasing $z$ to the maximum at $z\sim2.8$ in
both Scenarios B and C, and then decreases to almost zero at the formation redshift
$z_f$. At $z=1$, the number density $\phi^*$ for $L^*$ galaxies is 44\% and 88\%
higher than the present-day value for Scenarios B and C, respectively. Notice that the
Sdm/Irr/vB galaxies are star-forming galaxies, as suggested by their blue colors and
predicted by our population synthesis models. If we only focus on these galaxies, they
appear to be a disappearing or self-destructing population (cf. Babul \& Rees 1992;
Babul \& Ferguson 1996; Ferguson \& Babul 1998), but they are produced naturally by
the merger scenario, and are not needed to be ad hoc introduced into the models. In
particular, since the faint-end slope $\alpha$ also increases with $z$, there are even
more dwarf galaxies at high redshifts. Such a population seems to be qualitatively
consistent with the recent findings using Lyman-break selection techniques (see Steidel
\& Hamilton 1992, 1993; Steidel, Pettini, \& Hamilton 1995; Madau 1995) that a
substantial population of star-forming galaxies exists at $z \geq 3$ (Steidel et al.
1996; Steidel et al. 1998). We believe that such a population of Lyman-break galaxies
can be identified with the Sdm/Irr/vB galaxies suggested by our models.

Such an analysis, however, cannot be applied to the case of Scenario A. In Figure 7,
$\phi^*$ becomes negative when $z>3$, which is obviously unphysical. The reason
why $\phi$ becomes negative at high redshifts is that a flat universe has less volume at 
a given redshift than an open or $\Lambda$-dominated universe. Therefore, in order to 
account for the faint-end counts, the number evolution of early-type spirals for
Scenario A has to be larger than that for Scenarios B and C (See Table 3). On the other
hand, the number evolution of ellipticals for Scenario A is less than for Scenarios B and
C. As a result, the increase of the mass density for early-type spirals exceeds the
corresponding decrease for ellipticals. Therefore, if the model is subject to the constraint
of the mass density conservation, it will inevitably predict a negative $\phi^*$ for
Sdm/Irr/vB galaxies at high-redshifts. The implication is that the mass density
conservation suggested by Eq. [4] cannot be applied to the case of Scenario A, and
furthermore, in order to explain the faint data, the mass density must increase with
$z$.

The failure of the assumption of mass density conservation in Scenario A can be 
understood in several different ways, and its cosmological interpretation is also not
unique. There may be a disappearing/self-destructing population of star-forming
galaxies at high redshift which has no present-day counterpart. Another possibility
is that galaxies have been losing mass during their evolution, in particular, during the
period of merging between galaxies, due to supernovae which enable the gas to escape
from the potential well of dark matter halos. The situations may be very complicated.

Regardless of the difficulty in explaining the physical mechanism behind the
non-conservation of the mass density for galaxies, we can follow the same ad hoc
prescription as that for early-type spirals to adjust the parameters ($Q_0$, $Q_2$,
$\gamma_1$, and $\gamma_2$) of the models to match the data, without
discriminating between the two possibilities discussed above. The values of these
parameters are also listed in Table 3. The predicted number counts for the latest type
galaxies are shown in Panel (c) of Figure 5. It can be seen that the predictions can be
accepted within the uncertainties. In addition, the predictions for the overall population
(shown in Panel (d) of Figure 5) are also acceptable.

The morphological number counts in the near-infrared bands have also been provided
by NICMOS (Teplitz et al. 1998). Predictions of the NLE models are presented in Figure
8 to compare with the observed data in the $H_{1.6}$ band ($\lambda_{eff}$=
1.6$\mu$m). We can see that the models can reasonably reproduce the data.

\subsection{Number counts in $B$- and $K$- passbands}

We now examine whether the number evolution models can reproduce $B$- and 
$K$-bands number counts obtained from the majority of surveys so far. From Figure 
9 we can see that the models can reproduce better the number counts at bright 
magnitudes in both the $B$ ($B<21$) and the $K$ ($K<17$) bands, indicating that
the normalization (involving both $\phi^*_0$ and $L^*$, cf. Ellis 1997) is appropriate
using the adopted LFs. (Due to large uncertainties in the local LFs, the normalizations
are usually chosen to scale the model predictions to the observed number counts at a
relatively faint magnitude of $B \sim 19.0$. See Pozzetti et al. 1996 for a comprehensive
review). Furthermore, the NLE models can reproduce the counts at the faintest $B$ and
$K$ magnitudes for each of the three Scenarios.

\subsection{Redshift distribution}

The spectroscopic studies with the LRIS spectrograph on the Keck Telescope of two of
the Hawaii deep survey fields SSA 13 and SSA 22 (Cowie et al. 1996) and the CFRS
present two of the largest and deepest redshift samples so far, providing powerful tools
for understanding the evolution of galaxies.

In Fig.10-(a), (b), and (c), we show the predicted $z$-distributions in the
magnitude range of $22.5<B<24.0$ in the three Scenarios, to compare with measured
redshifts from Cowie et al. (1996), whose completeness is more than 85\%. We can see
that the predictions roughly match the observed data, and in particular, there are no
very high-$z$ tails. In Fig.10-(d), (e), and (f), we compare the NLE-model predicted 
$z$-distributions, limited in the magnitude range of $17.5<I_{AB}<22.5$, with the
CFRS sample (Lilly et al. 1995a; Le F\`evre et al. 1995; Lilly et al. 1995b; Hammer et
al. 1995; Crampton et al. 1995). Notice that the CFRS data are less deep than the Hawaii
sample, with the completeness $\sim$85\%. We can see that the predictions are
reasonably matched by the data, especially for Scenarios B and C. However, we should
also realize that all models overpredict galaxies with $z>1$. The overprediction may be
due to the $z$-unidentified galaxies, since the completeness of both of the two samples
is around 85\%, and, according to Pozzetti et al., such $z$-unidentified galaxies are most
probably at high $z$ (see Section 3.6 of Pozzetti et al. 1996).

\section{Summary and Conclusions}

Recent morphological data on faint galaxies are of great significance for us in
understanding the distant universe, and have shed new light on the long-standing
FBG problem. We have systematically investigated this question using HST data. In
previous work, we found that the number counts of ellipticals could be explained
successfully by PLE models only a scenario where ellipticals have {\it extremely}
short SFR timescales ($\tau_e=0.2$, 0, and 0.05 Gyr for flat, open and $\Lambda$
-dominated universes, respectively), i.e., they formed at high-$z$ (say $z_f=5$)
and have passively evolved since their formation (HZ98). Such a view, however,
would contradict other observations such as color distributions (HZ99a). Hence,
the evolution of galaxy number density seems to be required, and one of the main
purposes of this work is to study the nature of the number evolution.

PLE models are the starting point of our investigation. By modelling the number
counts of the three morphological classes in the HST $I_{814}$-passband, in 
combination with the analysis of redshift distributions in the three observed 
magnitude bins, $I_{814}<21.0$, $21.0<I_{814}<22.5$, and $22.5<I_{814}<25.0$, 
we find that PLE models can fit the number counts at the bright magnitudes of
$I_{814}<22.5$, indicating such PLE models could be valid up to $z\approx1$.
As a result, scenarios of strong number evolution at low redshifts such as that of
Kauffmann et al. (1996) would seem to be rejected. The results agree well with
those of Im et al. (1999), based on the morphologically divided redshift distribution
of faint galaxies, and those of Brinchmann et al. (1998) and Lilly et al. (1998), based
on the HST imaging of the CFRS and LDSS redshift surveys. In addition, our
conclusions are also consistent with those of Roukema \& Yoshii  (1993) and Woods
\& Fahlman (1997) based on analysis of the angular correlation function, and that of
Totani \& Yoshii (1998) based on the $V/V_{max}$ analysis.

We then constructed a set of phenomenological number-luminosity evolution models 
and by trial and error fixed a set of model parameters to explain the data such as
morphological number counts both in $I_{814}$- and in $H_{1.6}$-bands, as 
well as the counts in the $B$- and $K$-bands, and redshift distributions derived from
the Hawaii and the CFRS samples. The significant feature of these models is that the
number evolution is small at redshifts of $z<1$ and large at higher redshifts. We find
that the models can roughly reproduce the above-mentioned observations in each of the
cosmologies under consideration.

To simplify the modelling, we employ the constraint of conservation of the comoving
mass density of galaxies, in the form presented by Guiderdoni \& Rocca-Volmerange
(1991). We find that a population of late-type spiral/irregular galaxies and very blue
galaxies naturally emerge in these models. Such a population will peak at $z\approx2.8$.
We argue that such a population seems to be consistent with the recent findings using
Lyman-break photometric selection techniques. This conclusion, however, is only valid
in an open or $\Lambda$-dominated cosmological models. If the universe is flat, the
comoving number density of such a predicted population becomes negative at high
redshifts ($z>3$), and hence it is obviously physically unacceptable. Therefore, in order
to account for the faint data, such a population has to be introduced ad hoc into the
modelling. The failure of the mass conservation assumption in a flat universe model,
as discussed in Section 3.1, may imply the existence of a disappearing/self-destructing 
population of star-forming galaxies at high redshifts, whose present-day counterpart
is not detectable. Alternatively, galaxies may have been losing mass during their lifetimes,
due to supernova bursts, for example. If so, the evolutionary scenario for galaxies may
be extremely complicated.

We have not taken into account the evolution of metallicity with redshift, and details such
as the recycling of the residual gas ejected by dying stars. These would be future
improvements over our current simple prescriptions. Nevertheless, the models
faithfully reproduce the number counts and the redshift distributions, and these
results can be treated as heuristic guidelines for further investigations.

\section*{Acknowledgments}

PH should like to thank Mr. S. Charlot for providing us with their synthesis spectral
evolutionary models and helpful correspondences, and Dr. D. Crampton for making
the CFRS data available to us in electronic form. We acknowledge the valuable
discussion between Dr. B. F. Roukema. We thank Dr. Chris D. Impey for a thorough
reading of the first version of this paper, and for his constructive suggestions. We also
thank The State Key Laboratory of Science and Engineering Computing (LSEC) of
Academia Sinica for providing us with computer supports. This work is in part
supported by the National Natural Science Foundation of China.

\pagebreak

\input table1.tex
\input table2.tex

\input table3.tex

\pagebreak

\input fig1.tex
\input fig2.tex
\input fig3.tex
\input fig4.tex
\input fig5.tex
\input fig6.tex

\input fig7.tex
\input fig8.tex
\input fig9.tex
\input fig10.tex

\end{document}

%% file: table1.tex
\begin{deluxetable}{lcccccc}
\tablecolumns{7}
\tablewidth{30pc}
\tablecaption{$B$-band LF parameters for present-day galaxies.}
\tablehead{
  &\colhead{E/S0}&\colhead{Sab}&\colhead{Sbc}&\colhead{Scd}&\colhead{Sdm/Irr}& vB
}
\startdata
$\phi^*$  &  10.0      &        4.00   &      6.00   &  6.00         &  4.00        &        4.00   \nl
$M^*$     & $-21.0$ & $-20.80$ & $-20.80$ & $-20.80$ & $-20.70$ & $-20.50$ \nl
$\alpha$   & $-0.90$ & $-0.95$   & $-1.10$   & $-1.10$   & $-1.20$   & $-1.30$
\enddata
\tablenotetext{ }{Notes:}
\tablenotetext{ }{(1). $\phi^*$, $M^*$, and $\alpha$ are determined by trial and error,
 see text for details;}
\tablenotetext{ }{(2). $\phi^*$ in units of 10$^{-4}/$Mpc$^3$;}
\tablenotetext{ }{(3). vB galaxies, taken after Pozzetti et al. (1996),  are included to
  explain the bluest samples, e.g., CFRS. See text for details;}
\tablenotetext{ }{(4). The Hubble constant has been scaled to $h$=0.5.}
\end{deluxetable}

%% file: table2.tex
\begin{deluxetable}{lccccccccccc}
\tablecolumns{12}
\tablewidth{35pc}
\tablecaption{SFR parameter $\tau_e$ (measured in Gyr) and
modelled $B-I$ and $B-K$ colors for galaxies in the three
cosmological models under consideration.
}
\tablehead{
&\multicolumn{3}{c}{Scen. A}& &\multicolumn{3}{c}{Scen. B}&
&\multicolumn{3}{c}{Scen. C}\\
\cline{2-4} \cline{6-8} \cline{10-12}\\
Type & $\tau_e$ & $B-I$ & $B-K$ & &$\tau_e$ & $B-I$ & $B-K$ & &
$\tau_e$ & $B-I$ & $B-K$}
\startdata 
E/S0-I & 0.5 & 2.29 & 4.14 & & 1.0 & 2.34 & 4.17 & & 1.1 & 2.35 & 4.18 \nl
E/S0-II & 0.7 & 2.29 & 4.12 & & 1.4 & 2.33 & 4.16 & & 1.6 & 2.33 & 4.16 \nl
E/S0-III & 1.0 & 2.28 & 4.11 & & 2.1 & 2.30 & 4.11 & & 2.4 & 2.30 & 4.10 \nl
E/S0-IV & 1.5 & 2.25 & 4.07 & & 3.0 & 2.24 & 4.06 & & 3.1 & 2.24 & 4.07 \nl
E/S0-V & 2.1 & 2.21 & 4.00 & & 4.1 & 2.15 & 3.94 & & 4.3 & 2.15 & 3.94 \nl
Sab & 2.3 & 2.15 & 3.89 & & 3.3 & 2.17 & 3.93 & & 3.5 & 2.15 & 3.91 \nl
Sbc & 4.6 & 1.81 & 3.45 & & 6.3 & 1.84 & 3.51 & & 7.0 & 1.81 & 3.47 \nl
Scd     & 9.5 & 1.57 & 3.14 & & 13.5 & 1.60 & 3.19 & & 15.0 & 1.59 & 3.18 \nl
Sdm/Irr & $\infty$&1.37&2.87& &$\infty$&1.42 & 2.95 & 
                        &$\infty$& 1.43 & 2.96 \nl
vB & -   & 0.92 & 2.25  & & -    &0.92 & 2.25 & &   -  & 0.92 & 2.25
\enddata
\tablenotetext{}{Notes:}
\tablenotetext{}{(1). We choose a Scalo IMF for ellipticals, and a Salpeter IMF
 for the other types;}
\tablenotetext{}{(2). $\tau_e=\infty$ for Sdm/Irr galaxies denotes constant SFR;}
\tablenotetext{}{(3). The SFR of vB galaxies is not described by Eq. [1]. See text
  for detail.}
\end{deluxetable}

%% file: table3.tex
\begin{deluxetable}{lcccccc}
\tablecolumns{7}
\tablewidth{36pc}
\tablecaption{The phenomenological parameters for the number
evolution models in the three cosmological models under
consideration.}
\tablehead{\\
Parameter & E/S0 & Sab & Sbc & Scd & Sdm/Irr & vB\\ \\
\cline{1-7}\\
      &      &     & Scen. A &     &     &  \\
}
\startdata
$Q_0$      &  $-0.02$  &  0   &  0   &  0   &  1.0  &  1.0  \nl
$Q_2$      &  $-0.05$  & 0.03 & 0.03 & 0.03 & 0.03  & 0.03  \nl
$\gamma_1$ &    0    & 0.2  & 0.2  & 0.2  & 0.2   &  0.2  \nl
$\gamma_2$ &    0    & 0.5  & 0.5  & 0.5  & 0.5   &  0.5  \nl
\cline{1-7}\\
%
      &      &     & Scen. B &     &     &  \\ \\
\cline{1-7}
$Q_0$      &  $-0.05$  &  0   &  0   &  0   &  --  &  --  \nl
$Q_2$      &  $-0.15$  &  0   &  0   &  0   &  --  &  --  \nl
$\gamma_1$ &    0    & 0.2  & 0.2  & 0.2  & 0.2   &  0.2 \nl
$\gamma_2$ &    0    & 0.4  & 0.4  & 0.4  & 0.4   &  0.4 \nl
\cline{1-7}\\
      &      &     & Scen. C &     &     &  \\ \\
\cline{1-7}
$Q_0$      &  $-0.2$    &  0   &  0   &  0   &  --   &  --  \nl
$Q_2$      &  $-0.15$  &  0   &  0   &  0   &  --   &  --  \nl
$\gamma_1$ &    0    & 0.2  & 0.2  & 0.2  & 0.2   &  0.2 \nl
$\gamma_2$ &    0    & 0.4  & 0.4  & 0.4  & 0.4   &  0.4
\enddata
\tablenotetext{ }{Notes:}
\tablenotetext{ }{$Q_0$ and $Q_2$ for Sdm/Irr and vB galaxies
in Scenarios B and C are determined by Eq. [4].}
\end{deluxetable}

%% file: fig1.tex
%
\begin{figure*}[hbt]
\centerline{\psfig{figure=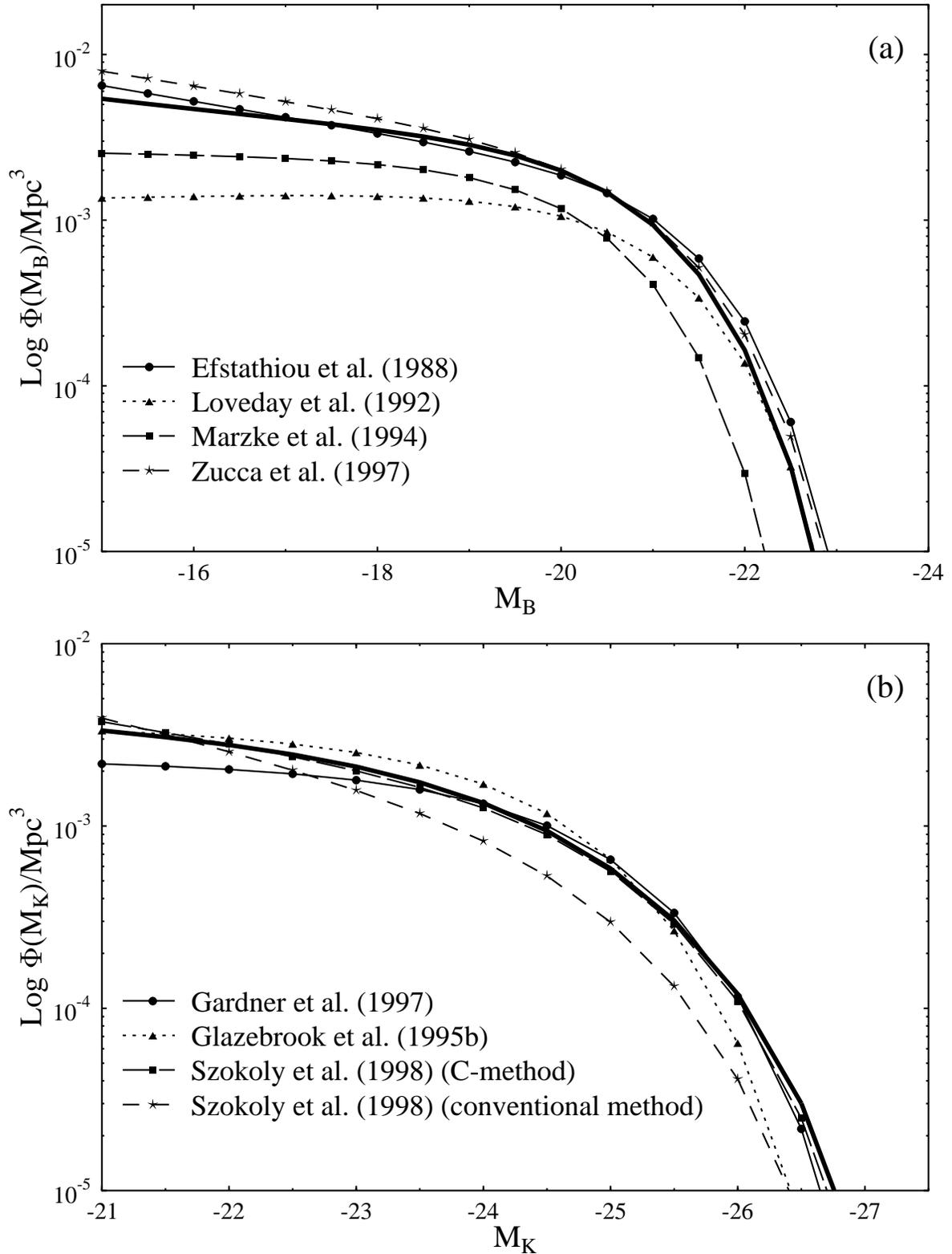,angle=0,width=16.0cm}}
\figcaption{
Present-day luminosity functions of galaxies. The adopted LFs are indicated by thick
lines, whose values correspond to those in Table 1. Here and hereafter, the capital
letters A, B and C refer to Scenarios A, B and C, respectively. Observational data were
obtained as follows. (a) Solid line with dots: Efstathiou, Ellis, \& Peterson (1988),
dotted line with triangles: Loveday et al. (1992), dashed line with squares: Marzke et
al. (1994), dashed line with stars: Zucca et al. (1997). (b) Solid line with dots: Gardner
et al. (1997), dotted line with triangles: Glazebrook et al. (1995b), dashed line with
squares: Szokoly et al. (1998; C-method), dashed line with stars: Szokoly et al. (1998;
conventional method). Notice that the characteristic density $\phi^*$ and the mixing
ratio between different galaxies of Efstathiou et al. (1988) LF are taken from Pozzetti et
al. (1996). Panels (a) and (b) are for $B$ and $K$ bands, respectively. The Hubble
constant has been scaled to $h=0.5$ for Scenario C.}
\end{figure*}

%% file: fig2.tex
%
\begin{figure*}[hbt]
\centerline{\psfig{figure=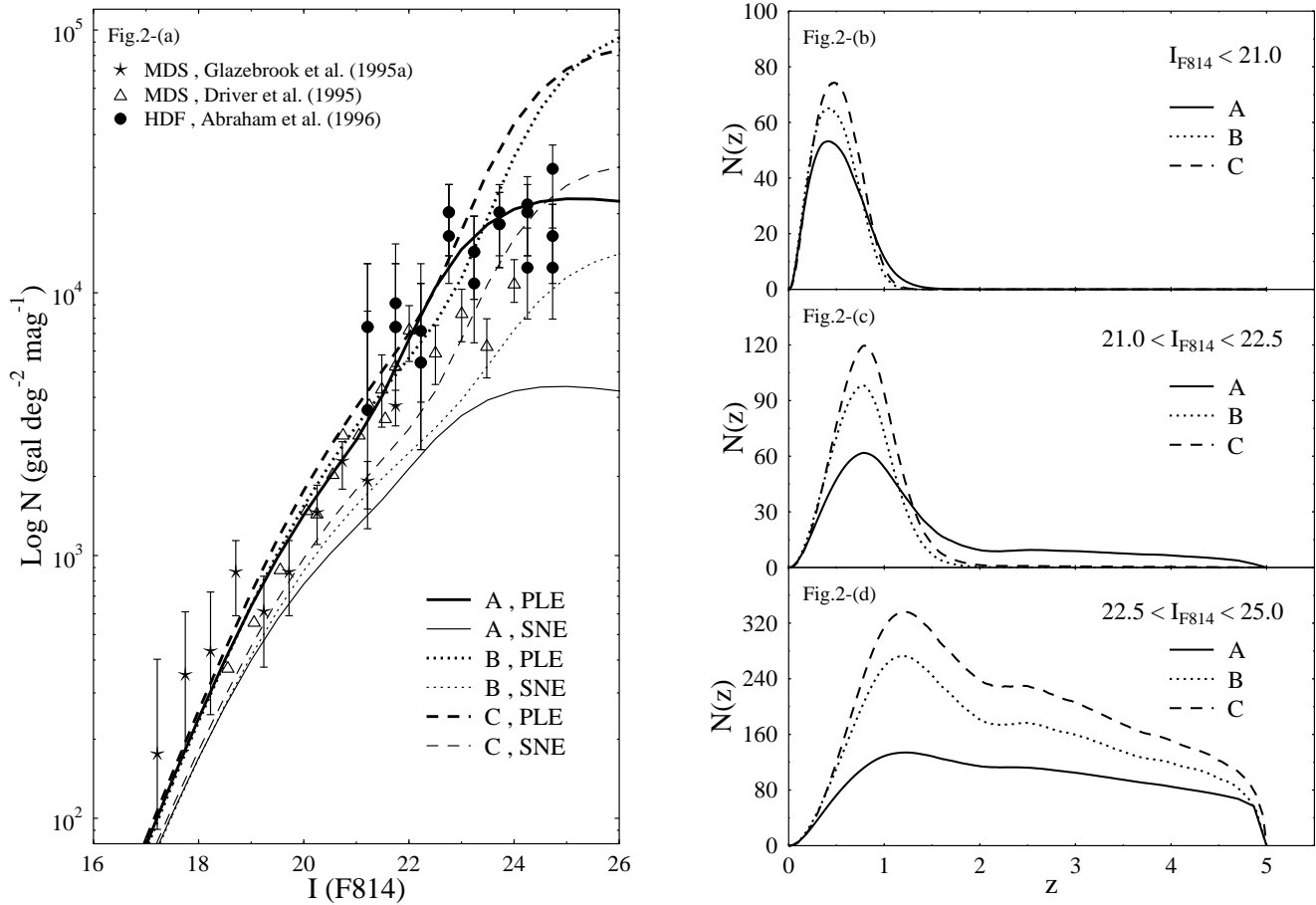,angle=0,width=18.0cm}}
\figcaption{
(a). Differential number counts for E/S0 galaxies as a function of apparent magnitude
in $I_{814}$ band. The sources of the observational data are exhibited in the figure and
these data are indicated by symbols. The model predictions are shown by lines. `PLE'
denotes pure luminosity evolution models, and `SNE', predictions of the strong number
evolution, based on the formalism of Kauffmann et al. (1996). See Section 2.1 for details.
Fig.2-(b), (c), and (d) are the $z$ distributions for E/S0 limited to three magnitude bins,
$I_{814}<21.0$, $21.0<I_{814}<22.5$ and $22.5<I_{814}<25.0$, respectively. Models
are indicated by lines, and the model predictions have been normalized in units of
deg$^{-2}$ with a redshift bin $\Delta z=0.01$.}
\end{figure*}

%% file: fig3.tex
%
\begin{figure*}[htb]
\centerline{\psfig{figure=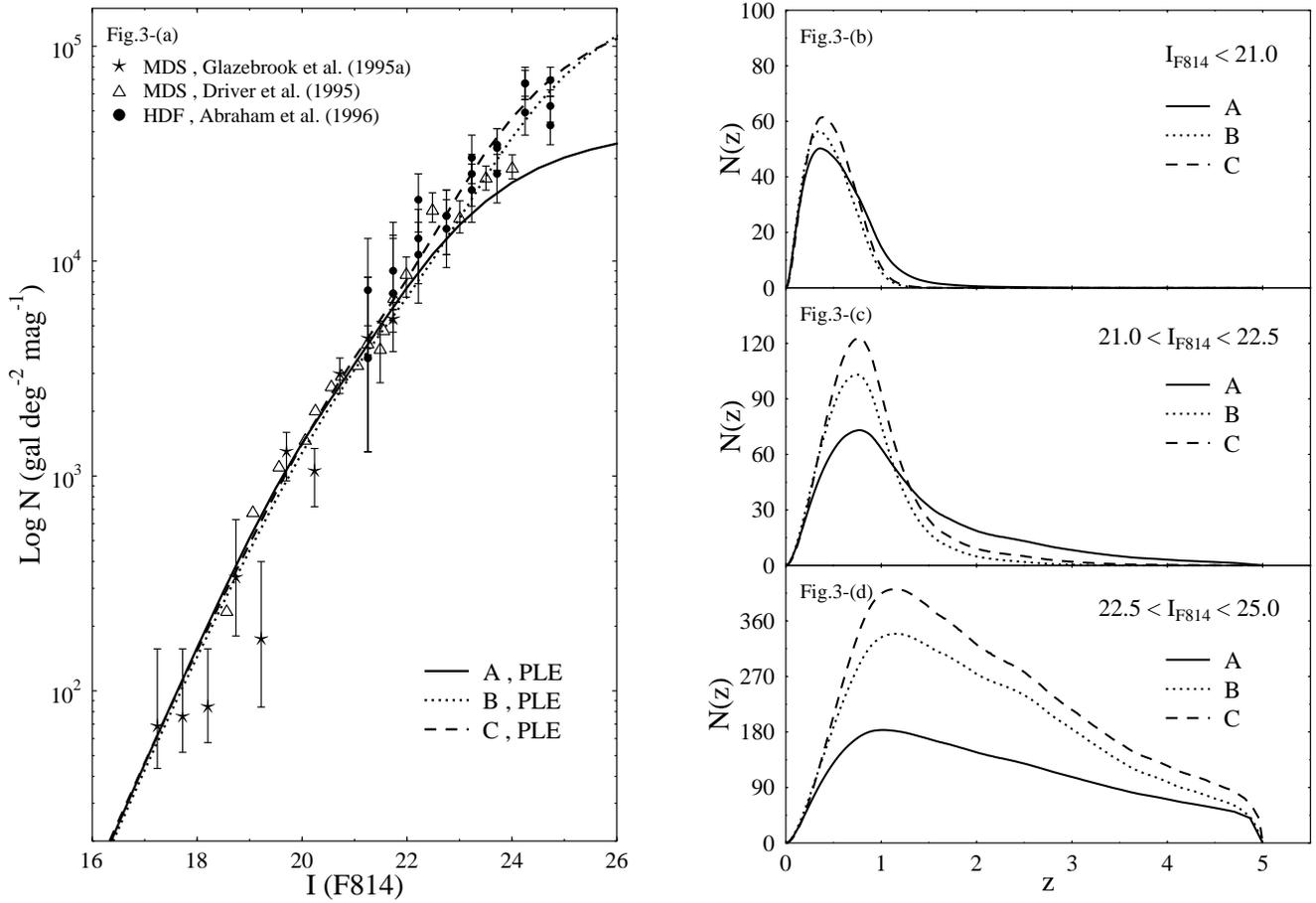,angle=0,width=18.0cm}}
\figcaption{
The same as Figure 2, but for early-type spiral galaxies (Sab+Sbc).
}
\end{figure*}

%% file: fig4.tex
%
\begin{figure*}[htb]
\centerline{\psfig{figure=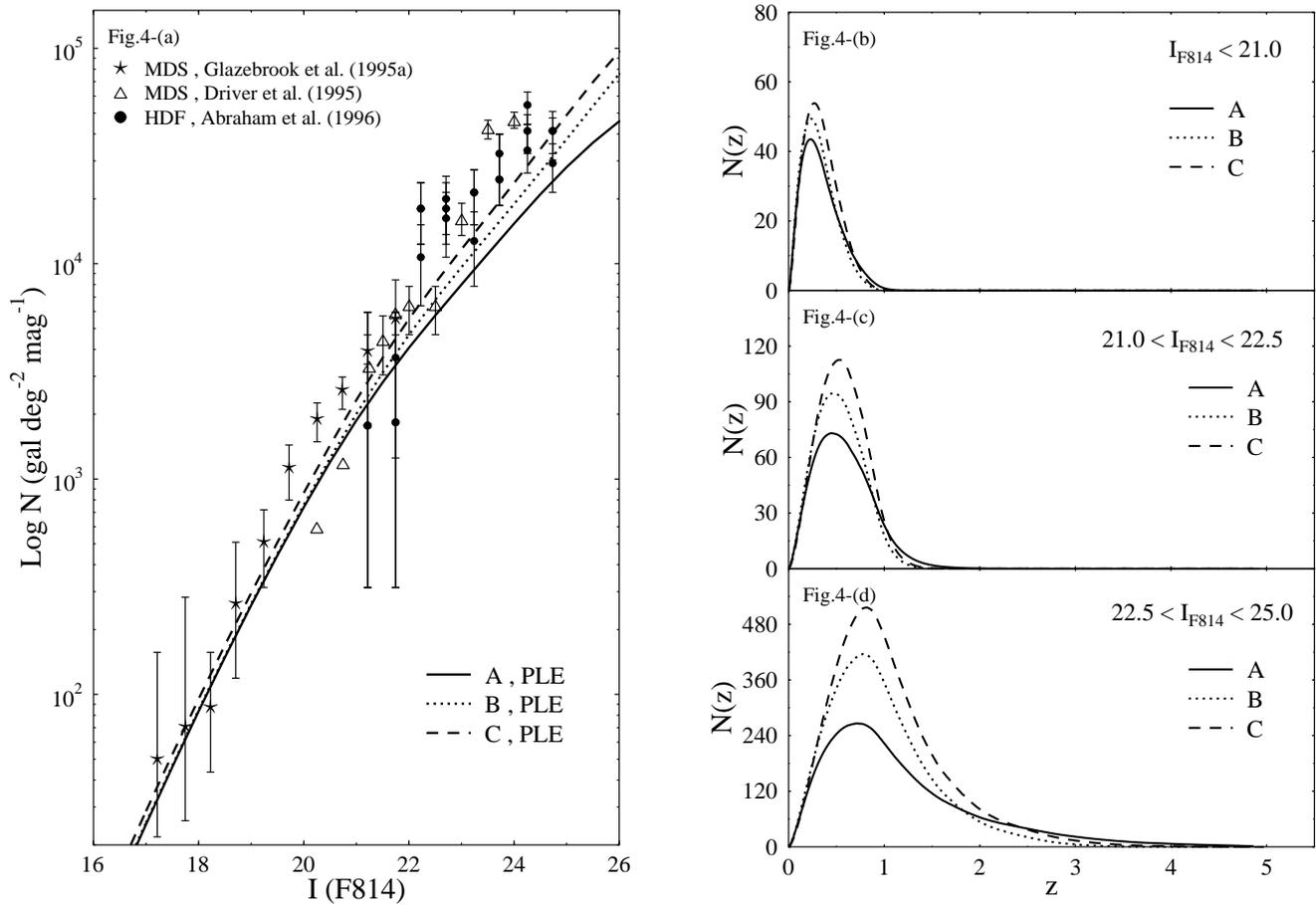,angle=0,width=18.0cm}}
\figcaption{
The same as Figure 2 and 3, but for late-type spiral/irregular
(Scd+Sdm/Irr) and vB galaxies.
}
\end{figure*}

%% file: fig5.tex
%
\begin{figure*}
\centerline{\psfig{figure=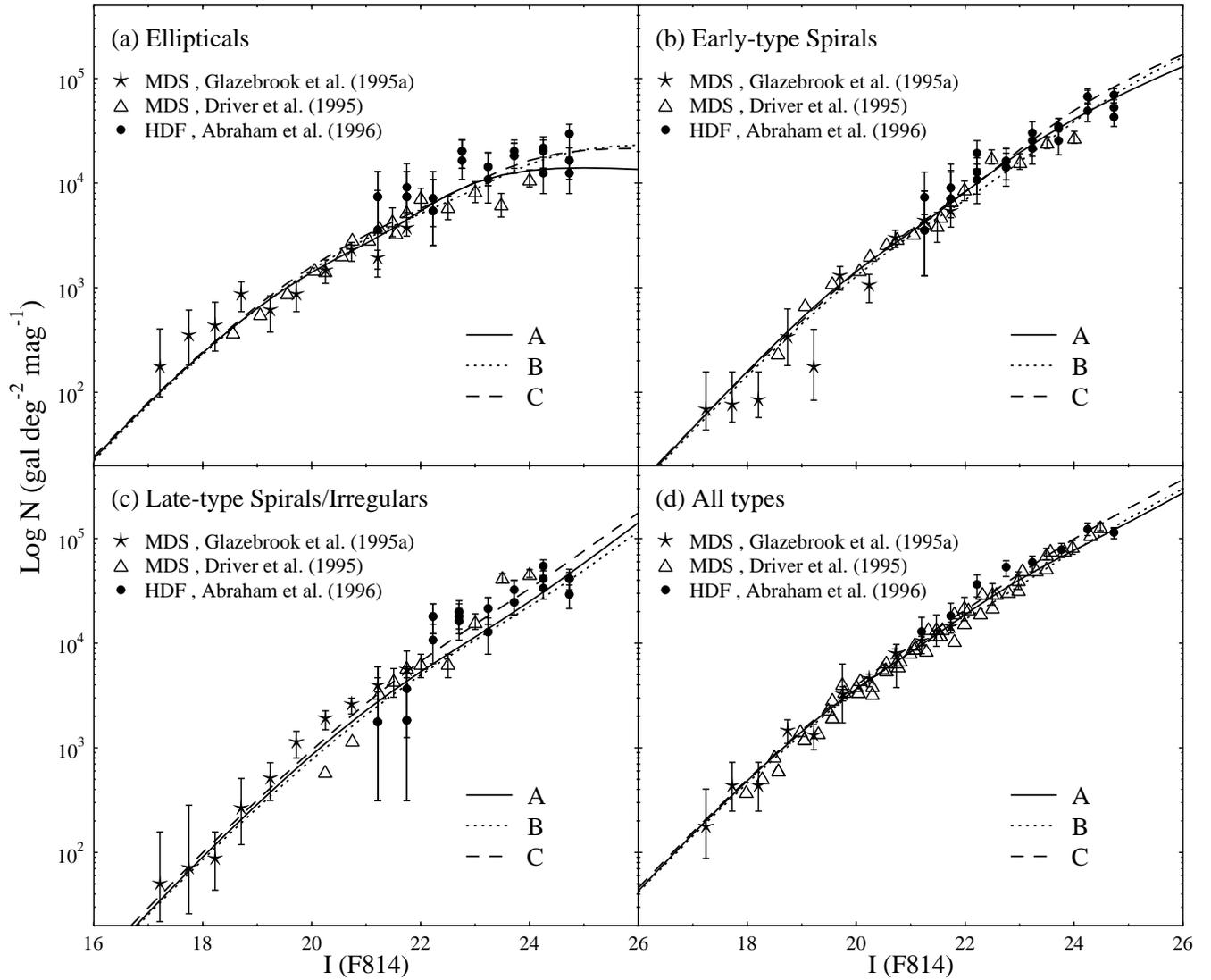,angle=0,width=18.0cm}}
\figcaption{
Differential number counts for galaxies as a function of apparent
magnitude in $I_{814}$ band. Panel (a), (b), (c) and (d) are for
ellipticals, early-type spirals, late-type spiral/irregular/vB galaxies
and the overall population, respectively. The model predictions are
shown by lines.}
\end{figure*}

%% file: fig6.tex
%
\begin{figure*}
\centerline{\psfig{figure=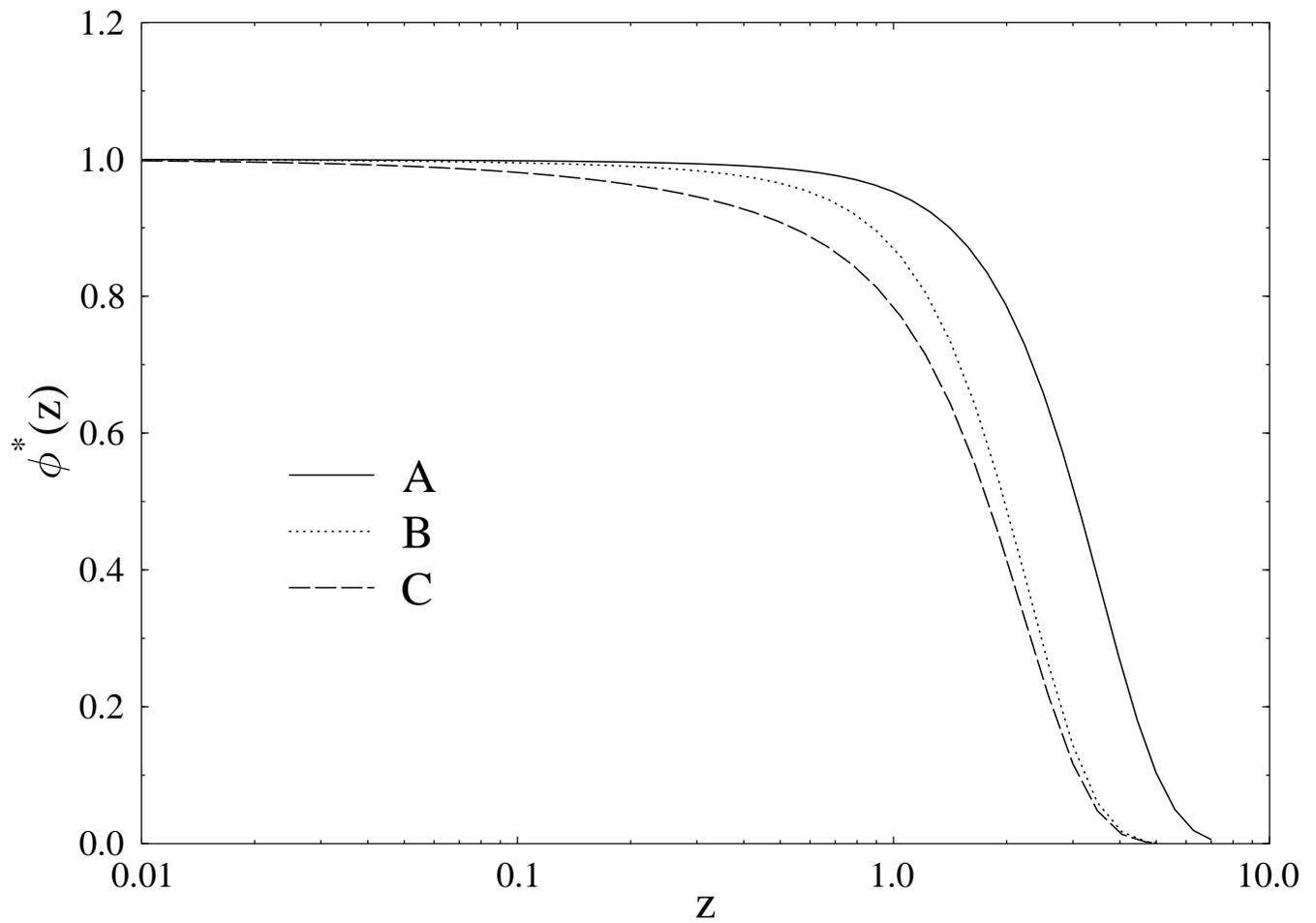,angle=0,width=18.0cm}}
\figcaption{
The evolution of the characteristic number density $\phi^*$ of E/S0
galaxies with respect to $z$. $\phi^*$ is normalized to unity at the
present day.}
\end{figure*}

%% file: fig7.tex
%
\begin{figure*}[htb]
\centerline{\psfig{figure=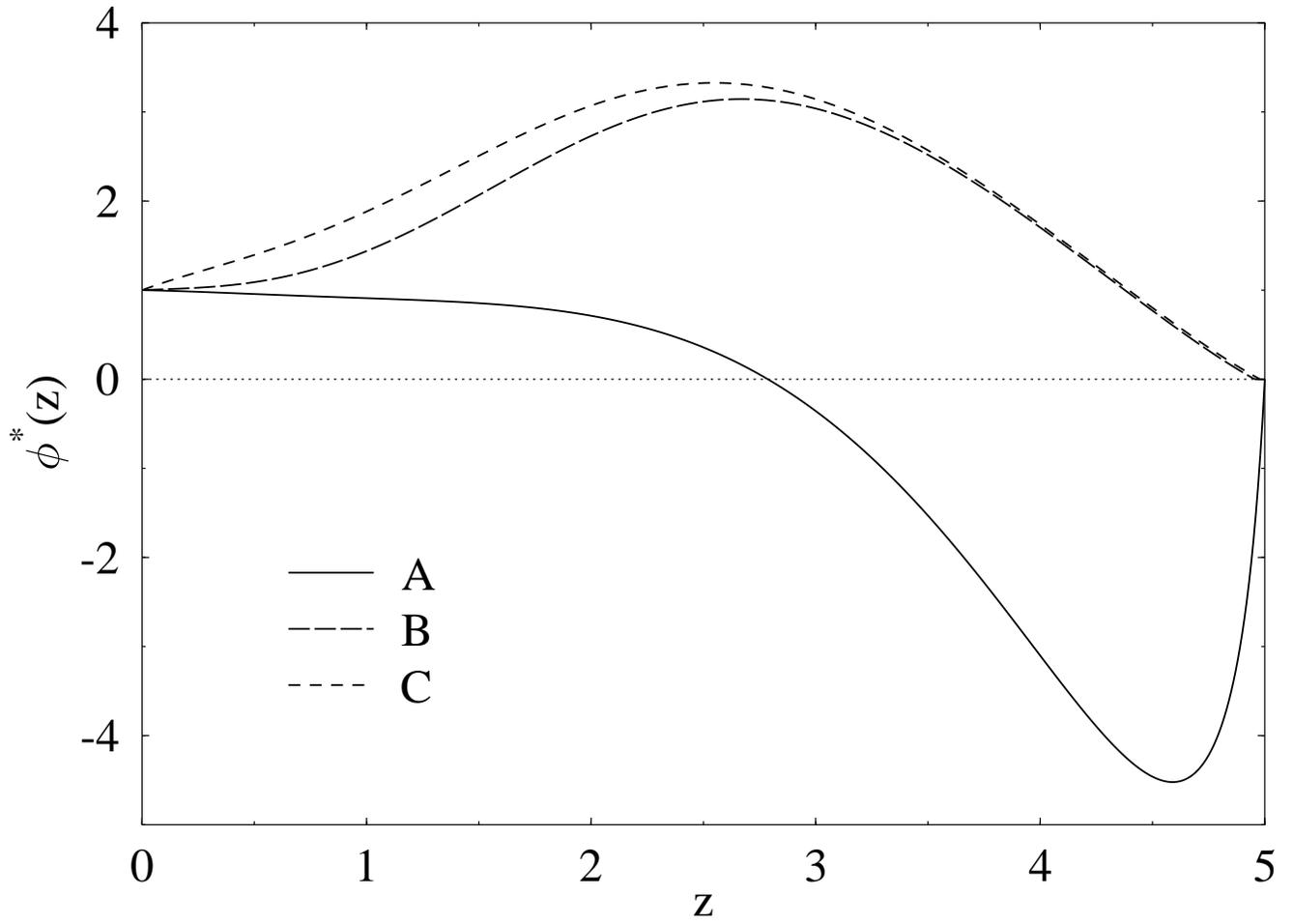,angle=0,width=18.0cm}}
\figcaption{
The evolution of the characteristic number density $\phi^*$ of Sdm/Irr
and vB galaxies with respect to $z$, constrained by the conservation of
the comoving mass density, Eq. [4]. $\phi^*$ is normalized to unity at the
present day.}
\end{figure*}

%% file: fig8.tex
%
\begin{figure*}
\centerline{\psfig{figure=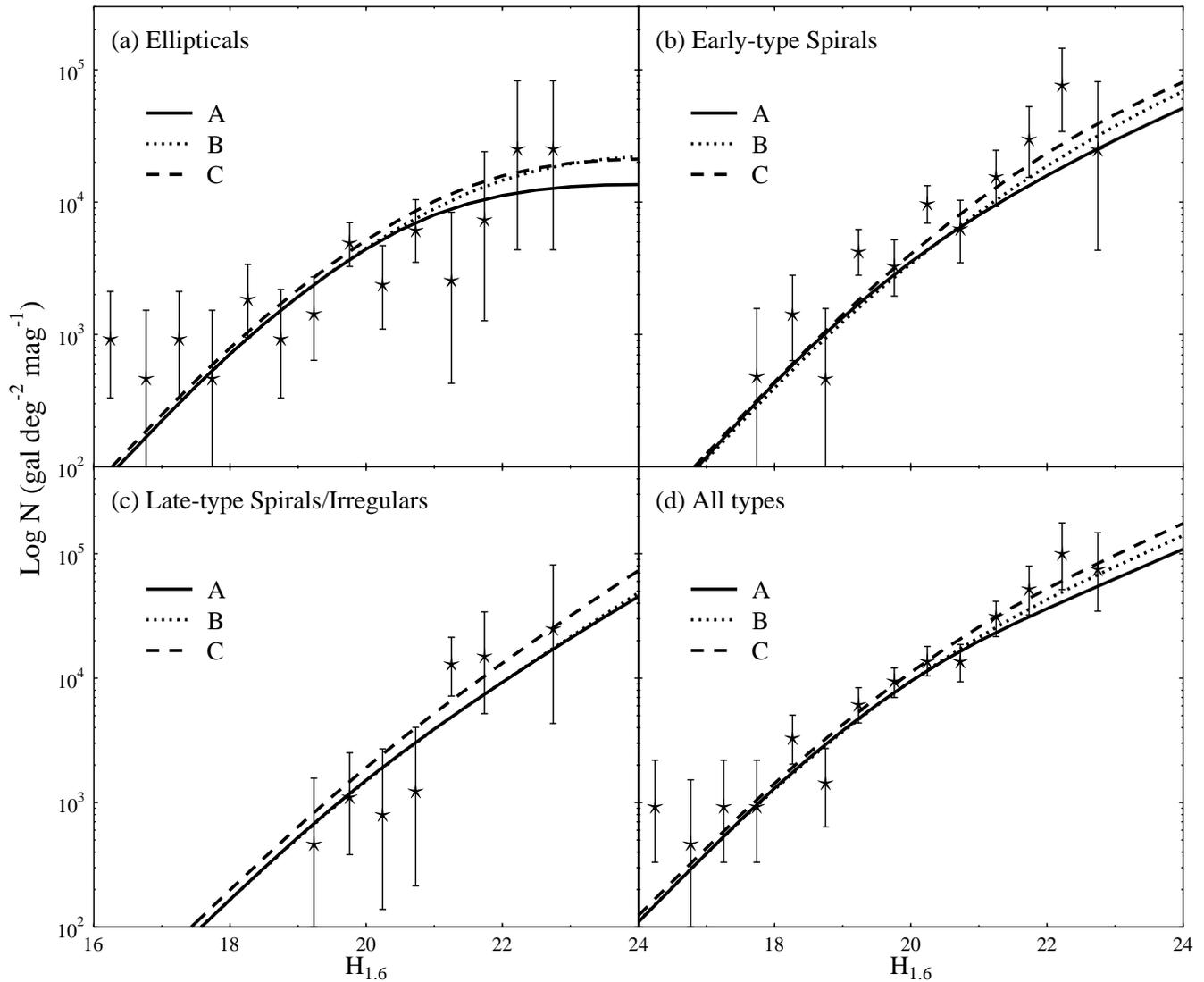,angle=0,width=18.0cm}}
\figcaption{
Differential number counts for galaxies as a function of apparent
magnitude in $H_{1.6}$ band. Panel (a), (b), (c) and (d) are for
ellipticals, early-type spirals, late-type spiral/irregular/vB galaxies
and the overall population, respectively. The data are taken from NICMOS
(Teplitz et al. 1998); the model predictions are shown by lines.}
\end{figure*}

%% file: fig9.tex
%
\begin{figure*}[htb]
\centerline{\psfig{figure=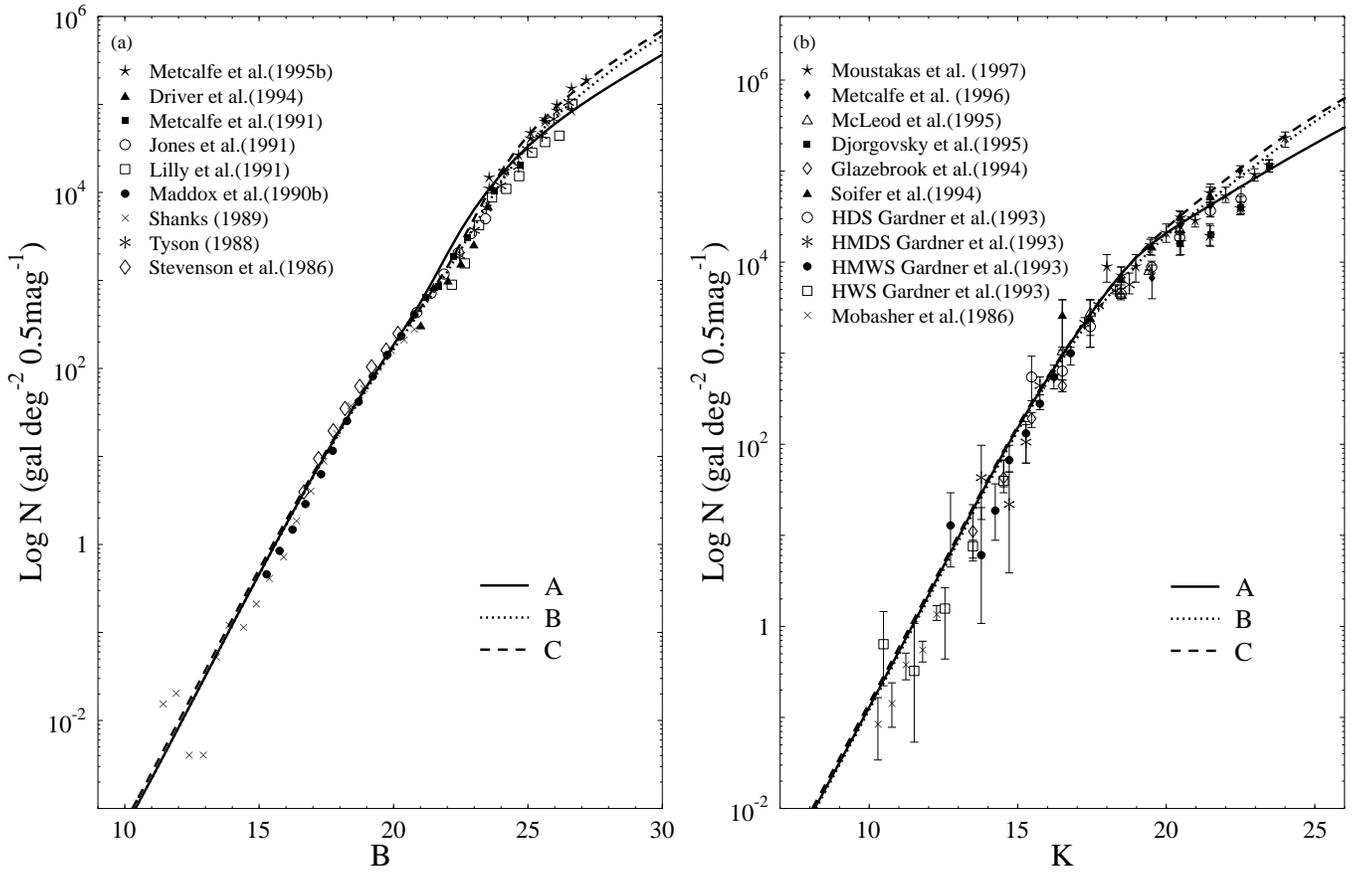,angle=0,width=18.0cm}}
\figcaption{
Differential number counts as a function of apparent magnitude in
$B$- and $K$- bands. The sources of observational data are exhibited
in the figure. Model predictions are shown by lines.
}
\end{figure*}

%% file: fig10.tex
\newpage
\begin{figure*}[htb]
\centerline{\psfig{figure=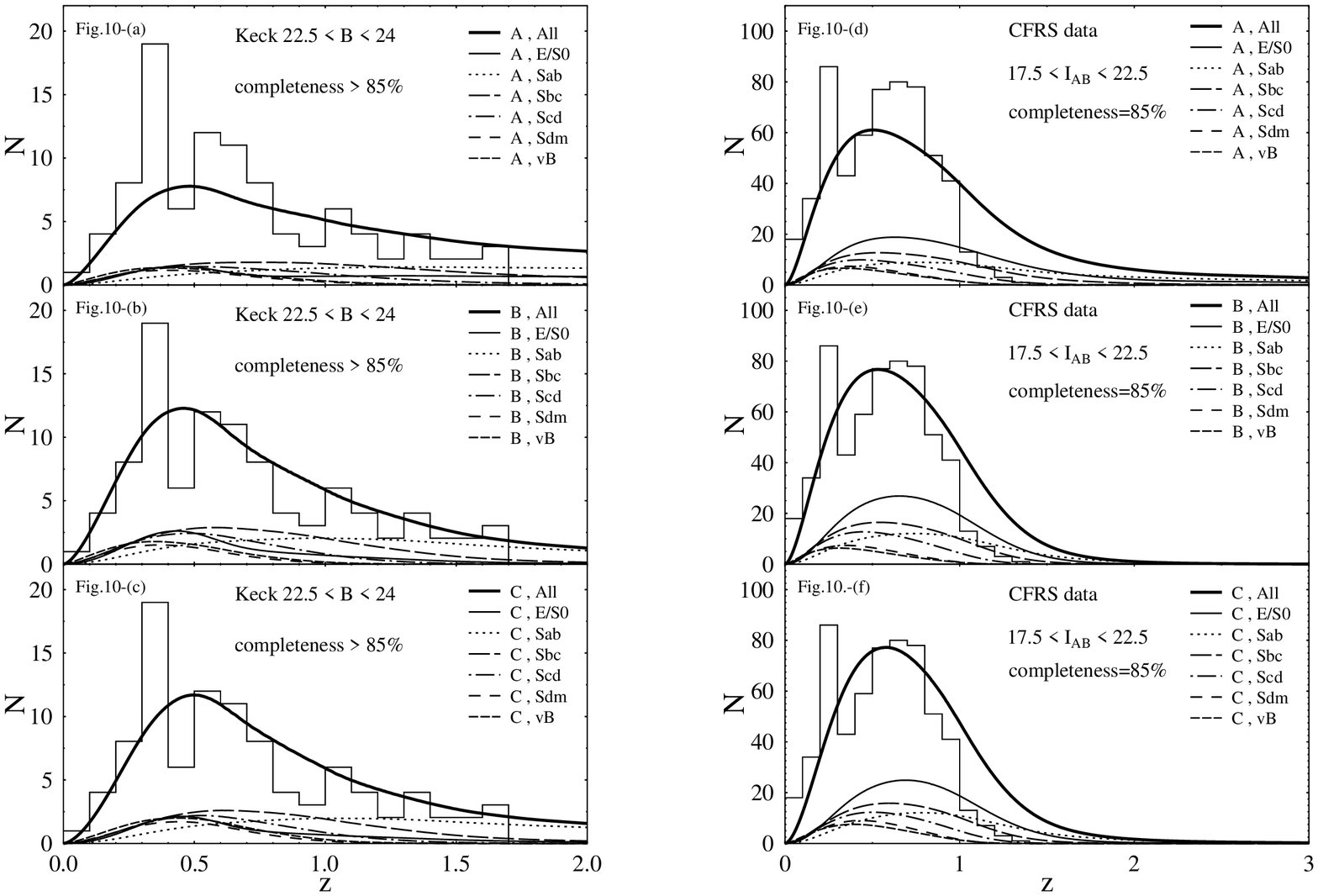,angle=0,width=18.0cm}}
\figcaption{
The $z$ distributions for galaxies. Fig.10-(a), (b), and (c) are limited
in the magnitude range of $22.5<B<24.0$, and for Scenario A, B, and C,
respectively. Fig.10-(d), (e), and (f) are corresponding to (a), (b),
and (c), respectively, but are limited in $17.5<I_{AB}<22.5$. The
sources of the observed data are indicated in the figure,
and they are shown by histograms. Model predictions are indicated by lines,
which have been normalized to give the total number of both $z$-identified
and $z$-unidentified objects.
}
\end{figure*}

%% file: mgr.bbl
\begin{references}

\reference{ } Abraham, R.G., Tanvir, N.R., Santiago, B.X., Ellis, R.S.,
   Glazebrook, K., and van den Bergh, S.: 1996, {\it Mon. Not. R. Astron. Soc.}
   {\bf279}, L47
\reference{ } Babul, A., and Rees, M.J.: 1992, {\it Mon. Not. R. Astron. Soc.}
   {\bf255}, 346
\reference{ } Babul, A., and Ferguson, H.C.: 1996, {\it Astrophys. J.} {\bf458}, 100
\reference{ } Binggeli, B., Sandage, A., and Tammann, G.A.: 1988,
  {\it Annu. Rev. Astron. Astrophys.} {\bf26}, 389 
\reference{ } Bower, R., Lucey, J.R., and Ellis, R.S.: 1992, {\it Mon. Not. R. 
  Astron. Soc.} {\bf254}, 601
\reference{ } Brinchmann, J. et al.: 1998, {\it Astrophys. J.} {\bf499}, 112
\reference{ } Broadhurst, T.J., Ellis, R.S., and Shanks, T.: 1988, 
   {\it Mon. Not. R. Astron. Soc.} {\bf235}, 827
\reference{ } Broadhurst, T.J., Ellis, R.S., and Glazebrook, K.: 1992, {\it Nature}
   {\bf355}, 55
\reference{ } Bruzual, A.G., and Kron, R.G.: 1980, {\it Astrophys. J.} {\bf241}, 25
\reference{ } Bruzual, A.G.: 1983, {\it Astrophys. J.} {\bf273}, 105
\reference{ } Campos, A., and Shanks, T.: 1997, {\it Mon. Not. R. Astron. Soc.}
   {\bf291}, 383
\reference{ } Carlberg, R.G., and Charlot, S.: 1992, {\it Astrophys. J.} {\bf397}, 5
\reference{ } Charlot, S., and Silk, J.: 1994, {\it Astrophys. J.} {\bf432}, 453
\reference{ } Cole, S., Aragon-Salamanca, A., Frenk, C.S., Navarro, J.F., and
   Zepf, S.E.: 1994, {\it Mon. Not. R. Astron. Soc.} {\bf271}, 781
\reference{ } Col\'in, P., Schramm, D.N., and Peimbert, M.: 1994, {\it Astrophys. J.}
   {\bf426}, 459
\reference{ } Colless, M., Ellis, R.S., Taylor, K., and Hook, R.N.: 1990,
   {\it Mon. Not. R. Astron. Soc.} {\bf244}, 408
\reference{ } Colless, M., Ellis, R.S., Broadhurst, T.J., Taylor, K.,
   and Bruce, A.: 1993, {\it Mon. Not. R. Astron. Soc.} {\bf261}, 19
\reference{ } Cowie, L.L.: 1991, {\it Phys. Scripta.} {\bf{T36}}, 102
\reference{ } Cowie, L.L., Songaila, A., and Hu, E.M.: 1991, {\it Nature} {\bf354}, 460 
\reference{ } Cowie, L.L., Songaila, A., Hu, E.M., and Cohen, J.G.:
   1996, {\it Astron. J.} {\bf112}, 839
\reference{ } Crampton, D., Le F\`evre, O., Lilly, S.J., and Hammer, F.:
   1995, {\it Astrophys. J.} {\bf455}, 96
\reference{ } Disney, M.J.: 1976, {\it Nature} {\bf263}, 573
\reference{ } Djorgovski, S., Soifer, B.T., Pahre, M.A., Larkin, J.E.,
   Smith, J.D., and Neugebauer, G. et al.: 1995, {\it Astrophys. J.} {\bf438}, L13
\reference{ } Driver, S.P., Phillipps, S., Davies, J.I., Morgan, I.,
   and Disney, M.J.: 1994, {\it Mon. Not. R. Astron. Soc.} {\bf266}, 155
\reference{ } Driver, S.P., Windhorst, R.A., and Griffiths, R.E.: 1995, 
    {\it Astrophys. J.} {\bf453}, 48
\reference{ } Driver, S.P., Windhorst, R.A., Ostrander, E.J., Keel, W.C.,
    Griffiths, R.E., and Ratnatunga, K.U.: 1995, {\it Astrophys. J.} {\bf449}, L23
\reference{ } Efstathiou, G., Ellis, R.S., and Peterson, B.A.: 1988,
    {\it Mon. Not. R. Astron. Soc.} {\bf232}, 431
\reference{ } Ellis, R.S.: 1997, {\it Annu. Rev. Astron. Astrophys.} {\bf35}, 389
\reference{ } Ferguson, H., and Babul, A.: 1998, {\it Mon. Not. R. Astron. Soc.}
    {\bf296}, 585
\reference{ } Ferguson, H., and McGaugh, S.S.: 1995, {\it Astrophys. J.} {\bf440}, 470
\reference{ } Fritze-v.Alvensleben, U., and Gerhard, O.E.: 1994, {\it Astron. Astrophys.}
    {\bf285}, 751
\reference{ } Fukugita, M., Takahara, F., Yamashita, K., and Yoshii, Y.: 1990,
    {\it Astrophys. J.} {\bf361}, L1
\reference{ } Gardner, J., Cowie, L.L., and Wainscoat, R.: 1993, {\it Astrophys. J.}
    {\bf415}, L9
\reference{ } Gardner, J.P., Sharples, R.M., Frenk, C.S., and Carrasco, B.E.:
   1997, {\it Astrophys. J.} {\bf480}, L99
\reference{ } Glazebrook, K., Peacock, J.A., Collins, C.A., and Miller, L.:
   1994, {\it Mon. Not. R. Astron. Soc.} {\bf266}, 65
\reference{ } Glazebrook, K., Ellis, R.S., Santiago, B.X., and Griffiths, R.:
   1995a, {\it Mon. Not. R. Astron. Soc.} {\bf275}, L19
\reference{ } Glazebrook, K., Peacock, J.A., Miller, J.A., and Collins, C.A.:
   1995b, {\it Mon. Not. R. Astron. Soc.} {\bf275}, 169
\reference{ } Gronwall, C., and Koo, D.C.: 1995, {\it Astrophys. J.} {\bf440}, L1
\reference{ } Guiderdoni, B., and Rocca-Volmerange, B.: 1990, {\it Astron. Astrophys.}
   {\bf227}, 362
\reference{ } Guiderdoni, B., and Rocca-Volmerange, B.: 1991, {\it Astron. Astrophys.}
   {\bf252}, 435
\reference{ } Hammer, F., Crampton, D., Le F\`evre, O., and Lilly, S.J.:
   1995, {\it Astrophys. J.} {\bf455}, 88
\reference{ } He, P., and Zhang, Y.Z.: 1998, {\it Mon. Not. R. Astron. Soc.} 
   {\bf298}, 483 (HZ98)
\reference{ } He, P., and Zhang, Y.Z.: 1999a, {\it Astrophys. J.} {\bf511}, 574 (HZ99a)
\reference{ } He, P., and Zhang, Y.Z.: 1999b, {\it Commun. Theor. Phys. (Beijing, China)}
  {\bf32}, 155
\reference{ } Im, M., Griffiths, R.E., Naim, A., Ratnatunga, K.U.,
   Roche, N., Green, R.F., and Sarajedini, V.L.: 1999, {\it Astrophys. J.} {\bf510}, 82
\reference{ } Impey, C., and Bothun, G.: 1997, {\it Annu. Rev. Astron. Astrophys.}
  {\bf35}, 267
\reference{ } Jones, L.R., Fong, R., Shanks, T., Ellis, R.S.,
   and Peterson, B.A.: 1991, {\it Mon. Not. R. Astron. Soc.} {\bf249}, 481
\reference{ } Kauffmann, G., White, S.D.M., and Guiderdoni, B.: 1993, {\it Mon.
   Not. R. Astron. Soc.} {\bf264}, 201
\reference{ } Kauffmann, G., Guiderdoni, B., and White, S.D.M.: 1994,
   {\it Mon. Not. R. Astron. Soc.} {\bf267}, 981
\reference{ } Kauffmann, G., Charlot, S., and White, S.D.M.: 1996,
   {\it Mon. Not. R. Astron. Soc.} {\bf283}, L117
\reference{ } Koo, D.C.: 1981, PhD Thesis, Univ. of California, Berkeley
\reference{ } Koo, D.C.: 1985, {\it Astron. J.} {\bf90}, 418
\reference{ } Koo, D.C., and Kron, R.G.: 1992, {\it Annu. Rev. Astron. Astrophys.}
  {\bf30}, 613
\reference{ } Le F\`evre, O., Crampton, D., Lilly, S.J., Hammer, F.,
   and Tresse, L.: 1995, {\it Astrophys. J.} {\bf455}, 60
\reference{ } Lilly, S.J., Cowie, L.L., and Gardner, J.P.: 1991, {\it Astrophys. J.} {\bf369}, 79
\reference{ } Lilly, S.J., Le F\`evre, O., Crampton, D., Hammer, F.,
   and Tresse, L.: 1995a, {\it Astrophys. J.} {\bf455}, 50
\reference{ } Lilly, S.J., Hammer, F., Le F\`evre, O., and Crampton, D.:
   1995b, {\it Astrophys. J.} {\bf455}, 75
\reference{ } Lilly, S.J. et al.: 1998, {\it Astrophys. J.} {\bf500}, 75
\reference{ } Loveday, J., Peterson, B.A., Efstathiou, G., and Maddox, S.J.:
   1992, {\it Astrophys. J.} {\bf390}, 338
\reference{ } Madau, P.: 1995, {\it Astrophys. J.} {\bf441}, 18
\reference{ } Maddox, S.J., Efstathiou, G., Sutherland, W.J., and Loveday, J.:
   1990a, {\it Mon. Not. R. Astron. Soc.} {\bf243}, 692
\reference{ } Maddox, S.J., Sutherland, W.J., Efstathiou, G., Loveday, J.,
   and Peterson, B.A.: 1990b, {\it Mon. Not. R. Astron. Soc.} {\bf247}, 1p
\reference{ } Marzke, R.O., Geller, M.J., Huchra, J.P., and Corwin Jr., H.G.:
   1994, {\it Astron. J.} {\bf108}, 437
\reference{ } McLeod, B.A., Bernstein, G.M., Rieke, M.J., Tollestrup,
   E.V., and Fazio, G.G.: 1995, {\it Astrophys. J. Suppl.} {\bf96}, 117
\reference{ } Metcalfe, N., Shanks, T., Fong, R., and Jones, L.R.:
   1991, {\it Mon. Not. R. Astron. Soc.} {\bf249}, 498
\reference{ } Metcalfe, N., Fong, R., and Shanks, T.: 1995a, 
   {\it Mon. Not. R. Astron. Soc.} {\bf274}, 769
\reference{ } Metcalfe, N., Shanks, T., Fong, R., and Roche, N.: 1995b,
   {\it Mon. Not. R. Astron. Soc.} {\bf273}, 257
\reference{ } Metcalfe, N., Shanks, T., Campos, A., Fong, R., and Gardner, J.P.:
   1996, {\it Nature} {\bf383}, 236
\reference{ } Mobasher, B., Ellis, R.S., and Sharples, R.M.: 1986,
   {\it Mon. Not. R. Astron. Soc.} {\bf223}, 11
\reference{ } Moustakas, L.A., Davis, M., Graham, J.R., Silk, J.,
   Peterson, B.A., and Yoshii, Y.: 1997, {\it Astrophys. J.} {\bf475}, 445
\reference{ } Pozzetti, L., Bruzual, A.G., and Zamorani, G.: 1996,
  {\it Mon. Not. R. Astron. Soc.} {\bf281}, 953
\reference{ } Rocca-Volmerange, B., and Guiderdoni, B.: 1990,
  {\it Mon. Not. R. Astron. Soc.} {\bf247}, 166
\reference{ } Roukema, B.F., and Yoshii, Y.: 1993, {\it Astrophys. J.} {\bf418}, L1
\reference{ } Roukema, B.F., Peterson, B.A., Quinn, P.J., and Rocca-Volmerange, 
   B.: 1997, {\it Mon. Not. R. Astron. Soc.} {\bf292}, 835
\reference{ } Salpeter, E.E.: 1955, {\it Astrophys. J.} {\bf121}, 161
\reference{ } Scalo, J.M.: 1986, {\it Fundam. Cosmic Phys.} {\bf11}, 1
\reference{ } Schechter, P.: 1976, {\it Astrophys. J.} {\bf203}, 297
\reference{ } Schmidt, M.: 1968, {\it Astrophys. J.} {\bf151}, 393
\reference{ } Shanks, T.: 1989, in The Extra-Galactic Background Light,
   eds. S. C. Bowyer \& C. Leinert (Kluwer Academic Publishers), p.269
\reference{ } Soifer, B.T., Matthews, K., Djorgovski, S., Larkin, J.,
   Graham, J.R., Harrison, W., Jernigan, G., and Lin, S. et al.: 1994,
   {\it Astrophys. J.} {\bf420}, L1
\reference{ } Steidel, C.C., and Hamilton, D.: 1992, {\it Astron. J.} {\bf104}, 941
\reference{ } Steidel, C.C., and Hamilton, D.: 1993, {\it Astron. J.} {\bf105}, 2017
\reference{ } Steidel, C.C., Pettini, M., and Hamilton, D.: 1995,
   {\it Astron. J.} {\bf110}, 2519
\reference{ } Steidel, C.C., Giavalisco, M., Pettini, M., Dickinson, M.,
   and Adelberger, K.L.: 1996, {\it Astrophys. J.} {\bf462}, L17
\reference{ } Steidel, C.C., Adelberger, K.L., Dickinson, M., Giavalisco, 
   M., Pettini, M., and Kellogg, M.: 1998, {\it Astrophys. J.} {\bf492}, 428
\reference{ } Stevenson, P.R.F., Shanks, T., and Fong, R.: 1986, in Spectral 
   Evolution of Galaxies., eds. C. Chiosi \& A. Renzini (Reidel, Dordrecht), p.439
\reference{ } Szokoly, G.P., Subbarau, M.U., Connolly, A.J., and Mobasher, B.:
   1998, {\it Astrophys. J.} {\bf492}, 452
\reference{ } Tinsley, B.M.: 1980, {\it Astrophys. J.} {\bf241}, 41
\reference{ } Teplitz, H.I., Gardner, J.P., Malumuth, E.M., and Heap, S.R.:
   1998, {\it Astrophys. J.} {\bf507}, L17
\reference{ } Totani, T., and Yoshii, Y.: 1998, {\it Astrophys. J.} {\bf501}, L177
\reference{ } Tyson, J.A.: 1988, {\it Astron. J.} {\bf96}, 1
\reference{ } Wang, B.: 1991, {\it Astrophys. J.} {\bf383}, L37
\reference{ } Woods, D., and Fahlman, G.G.: 1997, {\it Astrophys. J.} {\bf490}, 11
\reference{ } Worthey, G.: 1994, {\it Astrophys. J. Suppl.}, {\bf95}, 107
\reference{ } Zepf, S.E.: 1997, {\it Nature} {\bf390}, 377
\reference{ } Zucca, E., Zamorani, G., Vettolani, G., Cappi, A., Merighi, R.,
   Mignoli, M., Stirpe, G.M., and MacGillivray, H. et al.: 1997, {\it Astron. Astrophys.}
   {\bf326}, 477
\reference{ } Zwicky, F.: 1957, in Morphological Astronomy, Springer-Verlag 
\end{references}
